\documentclass[10pt,a4paper]{article}
\usepackage{amsmath}
\usepackage{amsfonts}
\usepackage{amssymb}
\usepackage{amsthm}
\usepackage{placeins}
\usepackage{changepage}
\usepackage{graphicx}
\usepackage{xspace}
\usepackage{hyperref}
\usepackage{booktabs}
\usepackage{chicago}
\usepackage{url}

\newcommand{\E}{\ensuremath{\mathbb{E}}}

\newcommand{\MB}{\ensuremath{\text{MB}}}

\newcommand{\Pbb}{\ensuremath{\mathbb{P}}}
\newcommand{\Fca}{\ensuremath{{\mathcal{F}}}}

\newcommand{\Qbb}{\ensuremath{{\mathbb{Q}}}}
\newcommand{\QbbTz}{\ensuremath{{\mathbb{Q}}}}

\newcommand{\CVA}{\ensuremath{\text{CVA}}}

\newcommand{\xCCno}{\ensuremath{\text{XVA}_\text{Client}}}
\newcommand{\xCCwith}{\ensuremath{\text{XVA}_\text{Client}^\text{MB}}}
\newcommand{\FVA}{\ensuremath{\text{FVA}}}

\newcommand{\xVA}{\ensuremath{\text{XVA}}}

\newcommand{\LGD}{\ensuremath{{L_{GD}}}}

\newcommand{\xo}{\ensuremath{X_\omega}}
\newcommand{\xoc}{\ensuremath{X_{\omega,C}}}
\newcommand{\yo}{\ensuremath{Y_\omega}}
\newcommand{\yoc}{\ensuremath{Y_{\omega,C}}}

\newcommand{\xva}{XVA\xspace}
\newcommand{\mb}{Mandatory Break\xspace}
\newcommand{\mbs}{Mandatory Breaks\xspace}
\newcommand{\cvs}{Covid-19\xspace}
\newcommand{\res}{Reset\xspace}
\newcommand{\ress}{Resets\xspace}
\newcommand{\resg}{Restructuring\xspace}

\theoremstyle{definition}
\newtheorem{definition}{Definition}[section]

\begin{document}
\title{Client engineering of \xva in crisis and normality:\\Restructuring, \mbs and \ress}
\author{Chris Kenyon\footnote{Contact: chris.kenyon@mufgsecurities.com. This paper is a personal view and does not represent the views of MUFG Securities EMEA plc (“MUSE”).   This paper is not advice.  Certain information contained in this presentation has been obtained or derived from third party sources and such information is believed to be correct and reliable but has not been independently verified.  Furthermore the information may not be current due to, among other things, changes in the financial markets or economic environment.  No obligation is accepted to update any such information contained in this presentation.  MUSE shall not be liable in any manner whatsoever for any consequences or loss (including but not limited to any direct, indirect or consequential loss, loss of profits and damages) arising from any reliance on or usage of this presentation and accepts no legal responsibility to any party who directly or indirectly receives this material.}}
\date{27 September 2020 \vskip5mm Version 1.10}

\maketitle

\begin{abstract}
Crises challenge client \xva management when continuous collateralization is not possible because a derivative locks in the client credit level and the provider's funding level, on the trade date, for the life of the trade.  We price \xva reduction strategies from the client point of view comparing multiple trade strategies using \mbs or \resg, to modifications of a single trade using a \res. We analyze previous crises and recovery of CDS to inform our numerical examples.  In our numerical examples \ress can be twice as effective as \mb/\resg if there is no credit recovery.  When recovery is at least 1/3 of the credit shock then \mb/\resg can be more effective.

\end{abstract}

%\newpage

%\tableofcontents

\section{Introduction}

Crises challenge client \xva management when continuous collateralization is not possible because a derivative locks in the client credit level and the provider's funding level, on the trade date, for the life of the trade.  We price \xva reduction strategies from the client point of view comparing multiple trade strategies using \mbs, \resg, to modifications of a single trade using a \res.  Multiple trade strategies are inefficient when there is no credit change because later trades have \xva priced in without including the probability of client survival, because only surviving clients will enter into continuation trades.  Pricing from the client point of view is necessary because continuation trades in multiple trade strategies are invisible to the provider by definition.  This means that pricing must use risk-neutral measures, and real-world-conditional risk neutral measures.  We analyze previous crises and recovery to inform our numerical examples on CDS shock sizes, and how long it takes a firm's CDS to recover by how much.

We price from the client perspective so \Pbb\ measures are important.  All \Pbb\ measures are subjective as they depend on user-chosen criteria, e.g. the calibration, or back-testing setup.   Our approach is to provide a mix of \Pbb-measure information, and scenarios to allow clients to assess risks of alternatives, not to collapse this information within an expectation.  This is because clients do not hedge own-credit and provider-funding, so we do not want to pre-judge which scenario is most important to clients.  This also avoids the anchoring effect of giving a single number since clients are not hedgers, unlike banks..

For credit shocks and recovery we analyze a comprehensive CDS database (2002 -- 20020) and give the historical \Pbb\ measure of shock recovery against time from shock for different shock sizes, see Table 1.  This analysis defines the range of credit recovery and timing we use in results Tables 2 and 3.  A vaccine for  SARS-CoV-2 may be months \cite{krammer2020sars} away, and  recovery from previous economic shocks generally took six months to 2--3 years.  In the numerical examples we consider CVA on an interest rate swap (IRS).  The \Pbb-conditional \Qbb\ measure is less important than might be expected because continuation trades are done at-the-money (ATM) so changes in rates levels are largely factored out.  We address changes in rates volatility by scenario analysis in Table 4.  

Pricing of derivatives from the client point of view seems to be absent in the literature, probably because clients are assumed to be price takers.  However, as we  demonstrate, clients can chose which prices (instruments) they take and when, to achieve their objectives.  This moves their price taking decisions into the realm of multi-stage stochastic optimization \cite{birge2011introduction} for portfolios.  However, we are interested in a simpler setup.  Design of hedging strategies for clients is a typical service provided by banks and informed by joint assessment of scenarios and risks.  Derivative pricing taking into account non-financial institution actions is typical to capture prepayment in Mortgage Backed Securities \cite{sirignano2016deep}.  Similar considerations apply for pricing revolving credit facilities, but the published literature is almost non-existent.

The contributions of this paper are firstly to price \xva from the client point of view which enables comparison of multiple trade and single trade \xva reduction strategies.  We provide a precise characterization of the required probability spaces, and conditional probability spaces.  Secondly, we compare: restructuring; \mbs; and \ress.  Thirdly we provide a quantification of CDS shocks and recoveries from history to inform choices of strategies and timing.  Finally we give numerical examples to quantify trade-offs of different strategies.  \xva reduction strategies must be priced from the client point of view and this is almost unique in the \xva literature.

\section{Client pricing}

We first give definitions and contract examples using  \mb/\resg and \res, then the probability framework.  We price from a client shareholder value point of view, not from a firm value point of view.  That is, we assume the client has no interest in events after their own default.

\subsection{\mb/\resg and \res definitions and examples}

\begin{figure}
\centering
\includegraphics[trim=0 0 0 0,clip,width=0.89\textwidth]{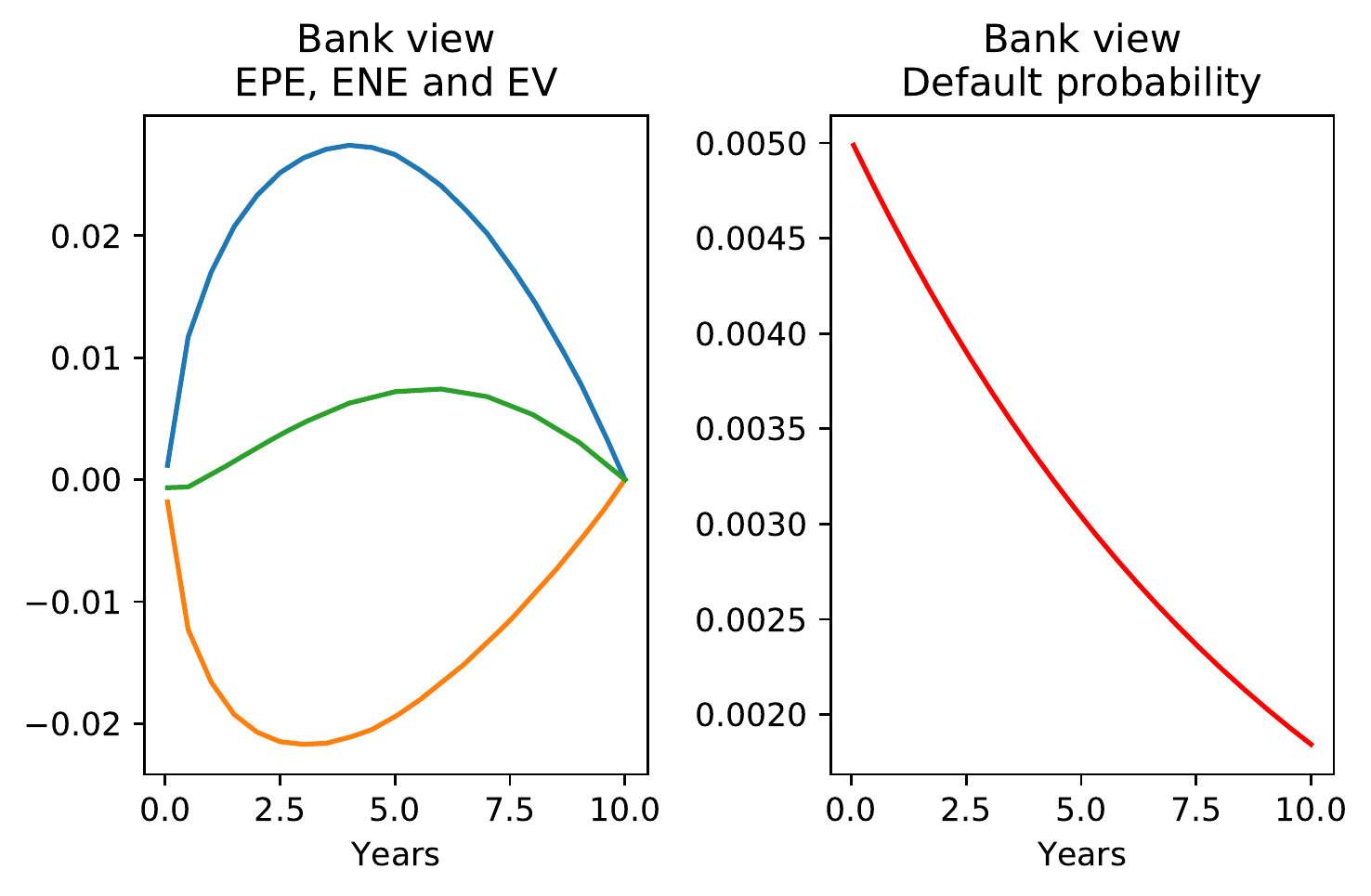}
\includegraphics[trim=0 0 0 0,clip,width=0.89\textwidth]{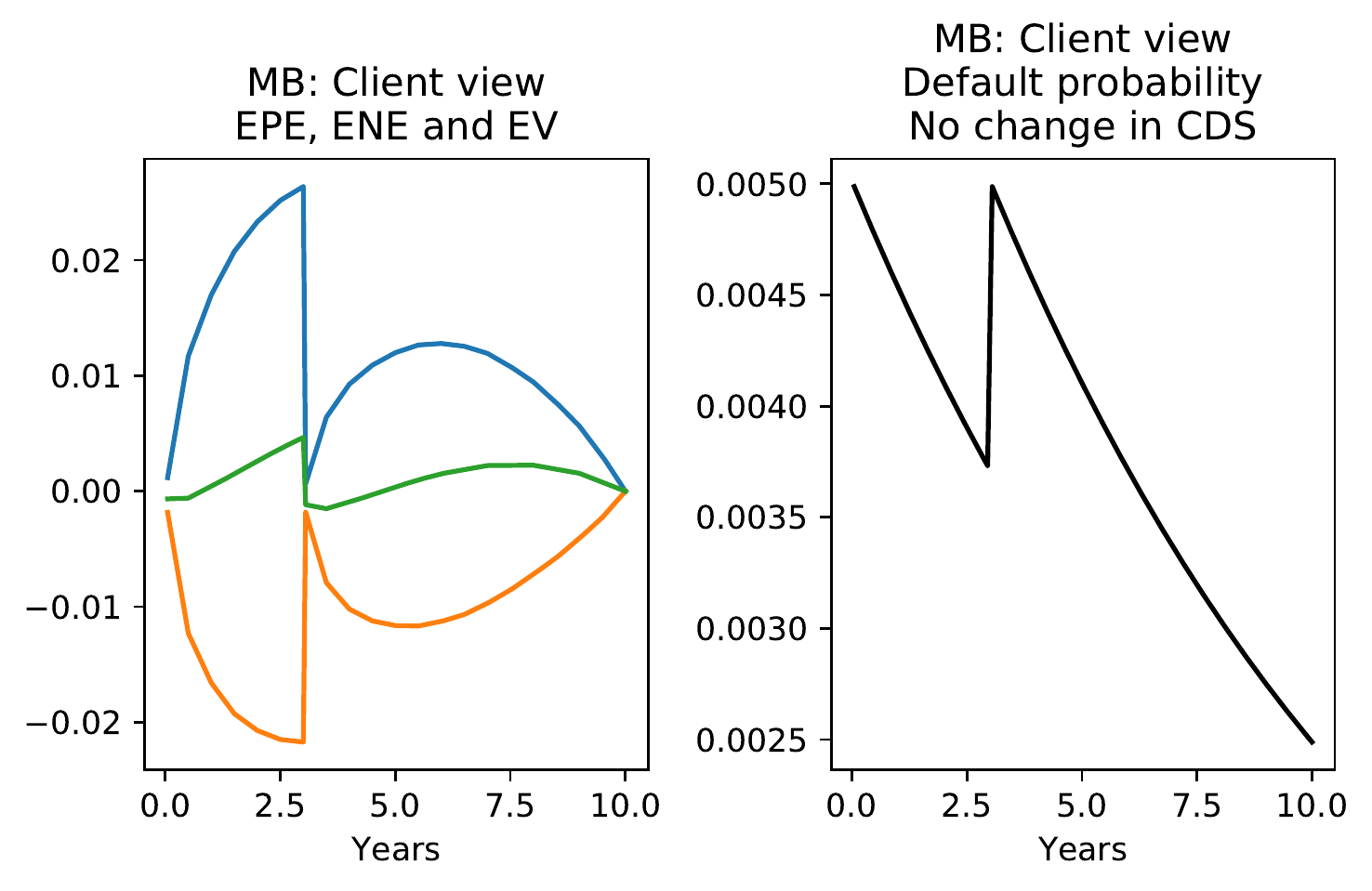}
\includegraphics[trim=0 0 0 0,clip,width=0.89\textwidth]{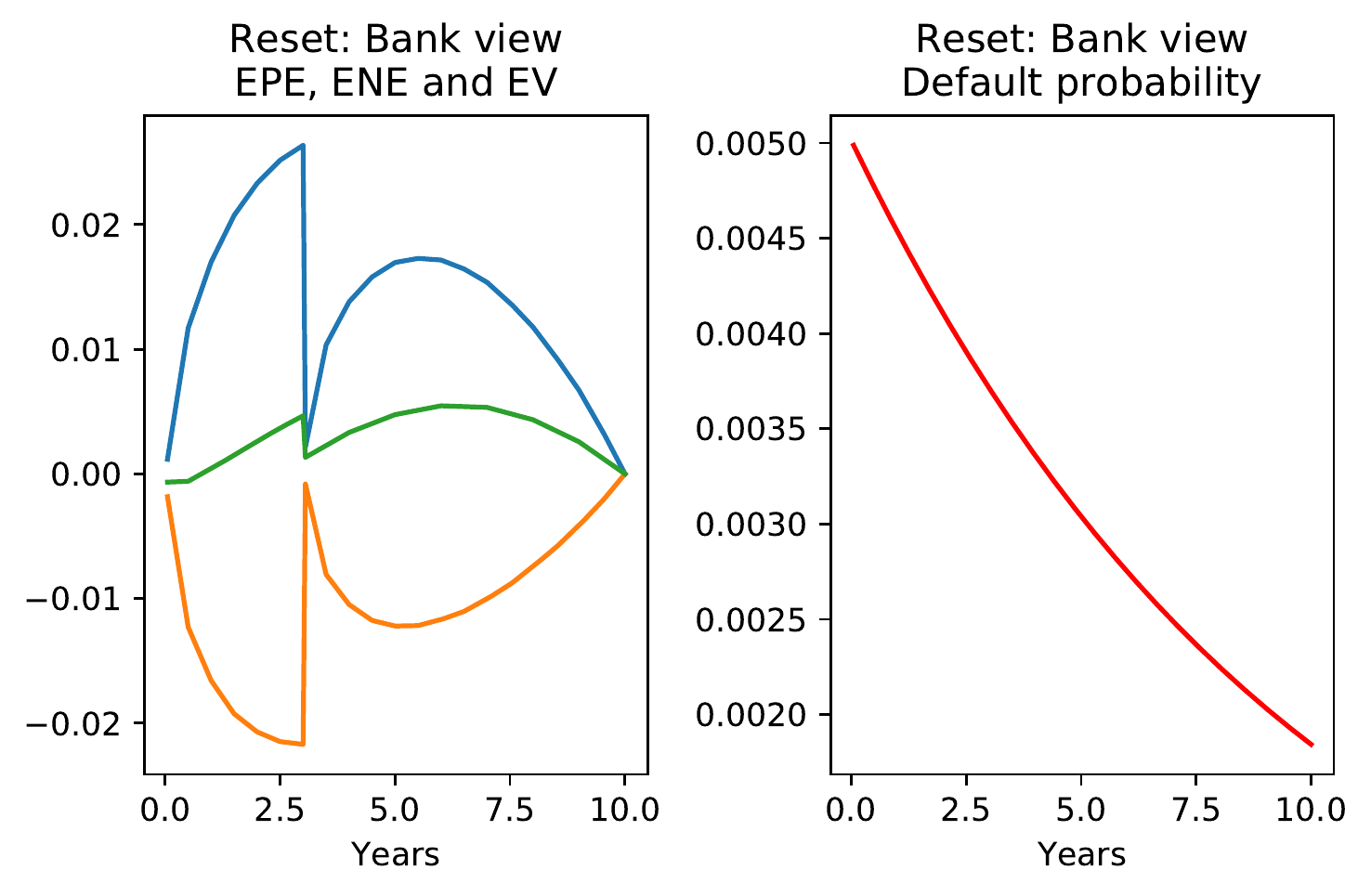}
\begin{adjustwidth}{-2.5cm}{-2.5cm}
\caption{TOP/LEFT: EPE, EV, and ENE profiles for 10y ATM EUR IRS with unit notional as of 2020-05-29.  TOP/RIGHT: default probability curve.  The curve gives the probability of default for the next 6m.
MID/LEFT: EPE, EV, and ENE profiles for same trade with a \mb after 3y and the profiles for the new 7y ATM IRS continuation trade assumed by the client that the bank uses from 3y to 10y.  MID/RIGHT: default probability curve used by the bank from $t_0$ to 3y, and the default probability curve assumed by the client that the bank uses from 3y to 10y.
BOTTOM/LEFT: EPE, EV, and ENE profiles for same trade with a \res after 3y.  BOTTOM/RIGHT: default probability curve used by the bank from $t_0$ to trade maturity.}
\label{f:examples}
\end{adjustwidth}
\end{figure}

\begin{definition}{\mb}
\begin{itemize}
\item A \mb is a legal agreement to end a derivative on the date specified, at the current market price, and is part of the termsheet.
\item The market price is defined as the price of the derivative ignoring default risk and funding costs.
\end{itemize}
\end{definition}
\res has the same effect as a \mb post-trade providing \xva rebates are available.  In \mb and \res the original contract stops and a new contract is entered for the remaining life of the original trade.  Since the new contract is only required by a surviving client the default probability resets as shown in Figure \ref{f:examples} MIDDLE, RIGHT.  The other key difference with a \res is that the credit and funding levels are also reset to whatever the current levels are at the time of the start of the new contract.  The profiles for \res and \mb after 3 years are slightly different because the \res is in the \Qbb-measure and the \mb continuation exposure is in the \Pbb-conditional-\Qbb-measure where we have picked the same-as-now future \Pbb\ measure.  Section \ref{ss:prob} provides a rigorous setup.

\begin{definition}{\res}
\begin{itemize}
\item A \res is a legal agreement to change some aspect of the trade on the date specified such that the NPV becomes zero, and payment of the NPV difference at the current market price, and is part of the termsheet.
\item The market price is defined as the price of the derivative ignoring default risk and funding costs.
\end{itemize}
\end{definition}

A {\bf multiple trade strategy} occurs with \mb, because there is a second trade after the \mb.  This second trade we call the {\bf continuation trade}.  This is also true for restructuring.

Figure \ref{f:examples} shows the exposure and default probability profiles of  the vanilla trade (TOP), then the effects of a \mb/\resg (MIDDLE) and \res (BOTTOM).

\subsection{CVA and FVA}
\label{s:xva}

Client valuation of trades with \ress is the same as that of the provider, there are no uncertainties in the price of CVA and FVA.  

Client valuation of trades with \mb/\resg includes the continuation trade after the \mb.  The continuation trade could be with a different provider to the original trade, and must be estimated by the client.  The market will also have moved by the \mb date so the client also needs to estimate this effect.  With a crisis the client aims to put the \mb after the crisis so as not to lock in the crisis-level credit and funding risks for any longer than necessary.  

When clients use \resg, they wait and observe the market before acting.  Choosing to potentially restructure later needs to be included in the original assessment of \xva to compare strategies.  We assume equivalence with \mb here for simplicity, i.e. there is 100\%\ rebate available on demand for \xva.

Thus clients view \xva from a future conditional measure perspective for \mb/\resg, because they do not hedge their own default and they do not hedge their derivative provider's funding cost and they assume their own survival.  This requires the following probability development.

\subsection{Probability spaces and conditional probability spaces}
\label{ss:prob}

To handle client valuation in the \Pbb\ measure conditional on their survival we introduce the probability space  
\[
X = (\Omega,\Fca,\Pbb)
\]
on a set of events $\Omega$ with a filtration $\Fca(t)$ and corresponding probability measures $\Pbb(t)$.  The equivalent probability space with a risk-neutral measure is
\[
Y = (\Omega,\Fca,\Qbb)
\]
$\Pbb(t)$ are the physical measures from the point of view of $t_0$.    Given a \mb date $t_m $ and a set of events (path) up to $t_m$, $\omega \in \Fca(t_m)$, we define sets of conditional probability spaces  from $X$ as
\begin{align}
X_\omega = & \{ (\Omega_\omega, \Fca_\omega, \Pbb_\omega)\  | \ \omega\in\Fca(t_m) \}  \\
X_{\omega,C} = & \{  (\Omega_{\omega,C}, \Fca_{\omega,C}, \Pbb_{\omega,C}) \  |\  \omega\in\Fca(t_m) \ \text{and}\ \tau_C > t_m \}
\end{align}

$\tau_C$ is the default time of the counterparty.

$\Omega_\omega$ are all possible events, conditional on the set of events $\omega$ up to $t_m$.

$\Fca_\omega$ is the filtration $\Fca$  conditional on the set of events $\omega$ up to $t_m$.

$\Pbb_\omega(t)$ are the probability measures $\Pbb(t)$ for $t\ge t_m$, conditional on the set of events $\omega$ up to $t_m$.

Hence $X_\omega$ is the set of  all future probability spaces at $t_m$, indexed by the state of the world $\omega$ up to $t_m$, and $X_{\omega,C}$ is the set of all future probability spaces where the client  survived up to and including $t_m$.  This modifies $(\Omega_\omega, \Fca_\omega, \Pbb_\omega)$ to $(\Omega_{\omega,C}, \Fca_{\omega,C}, \Pbb_{\omega,C})$ by adding the additional conditioning.

\begin{figure}
\centering
\includegraphics[trim=0 0 0 0,clip,width=0.45\textwidth]{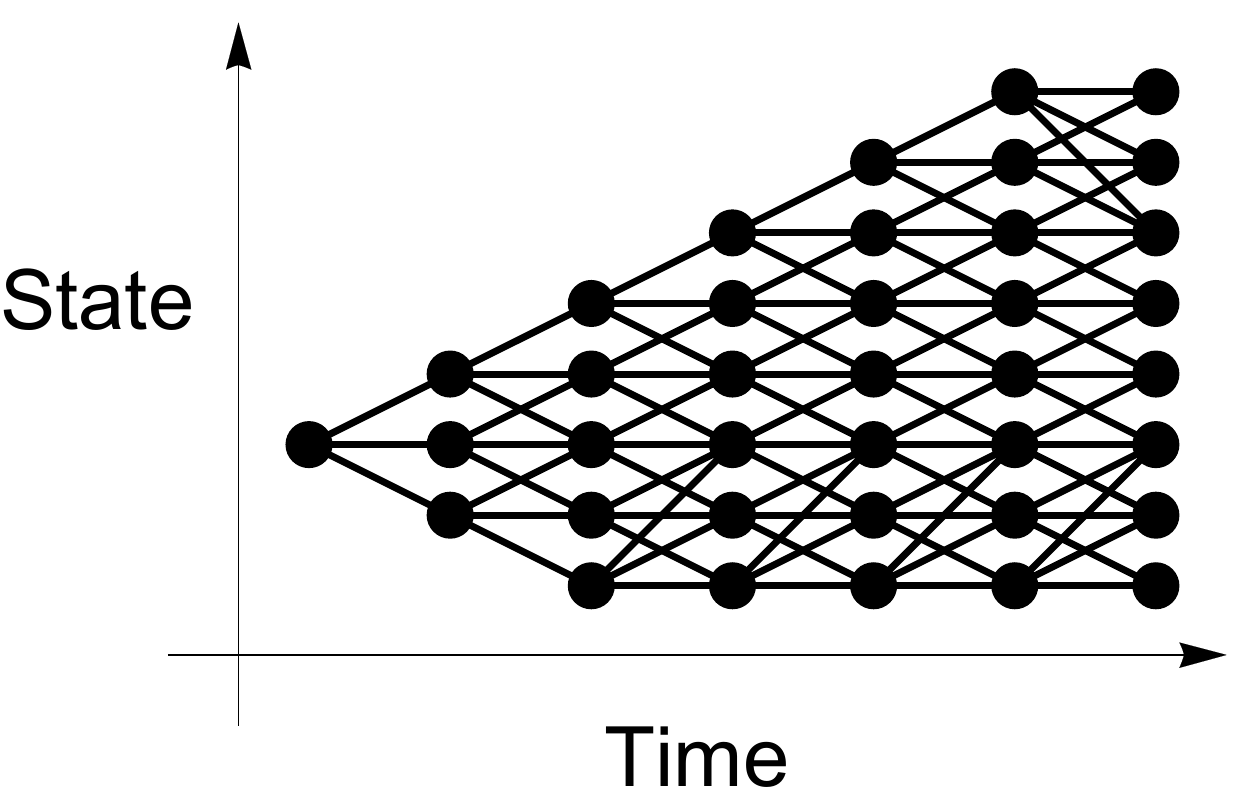}
\includegraphics[trim=0 0 0 0,clip,width=0.45\textwidth]{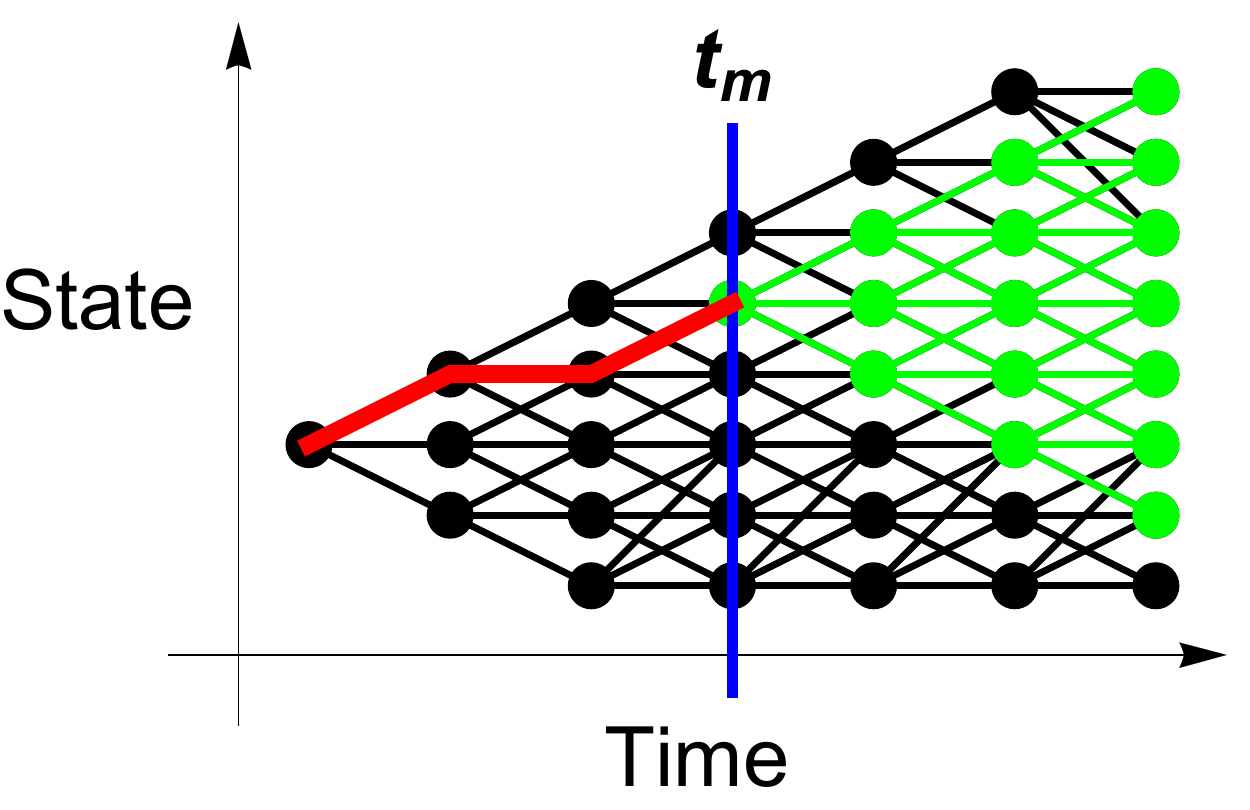}
\begin{adjustwidth}{-0cm}{-0cm}
\caption{Illustration of unconditional (LEFT), and conditional (RIGHT) probability spaces.  LEFT:  $X = (\Omega,\Fca,\Pbb)$, where $\Omega$ = dots, $\Fca$=lines.  RIGHT, $X_{\omega_a} = (\Omega_{\omega_a}, \Fca_{\omega_a}, \Pbb_{\omega_a})$ for a specific $\omega_a$ in $\Fca$ up to $t_m$ where $\omega_a$=red path; $\Omega_{\omega_a}$=green dots because only these are reachable from $\omega_a$;   $\Fca_{\omega_a}$=green lines, as these are the only futures reachable from $\omega_a$.  In $X_{\omega_a,C}$,   $\Omega_{\omega_a,C}$ will be the empty set if the client $C$ defaulted along the path $\omega_a$, otherwise $\Omega_{\omega_a,C} = \Omega_{\omega_a}$.}
\label{f:prob}
\end{adjustwidth}
\end{figure}

Figure \ref{f:prob} illustrates the probability spaces $X$ and $X_{\omega=\omega_a}$ for a specific $\omega_a$.  The vertical State axis indicates the multi-dimensional state of the world.  Lines indicate which states are reachable from each other.  We have chosen a recombining tree because it makes it easier to display $\Fca$ and  $\Fca_{\omega_a}$ .  In the context of the Figure, $X_{\omega,C}$ consists of those conditional probability spaces where the client does not default on the possible paths $\omega$ up to $t_m$.  So, for example, it may be that only some of the points at $t_m$ exist in $\cup_\omega \{\Omega_{\omega,C}\}$ considering all $\omega$ in $\Fca$ up to $t_m$.

Now for the probability spaces in \xo, or \xoc, we can create sets of equivalent risk-neutral probability spaces \yo, or \yoc, i.e.
\begin{align}
Y_\omega = &  \{ (\Omega_\omega, \Fca_\omega, \Qbb_\omega)\  | \ \omega\in\Fca(t_m) \} \\
Y_{\omega,C} = & \{ (\Omega_{\omega,C}, \Fca_{\omega,C}, \Qbb_{\omega,C}) \  |\  \omega\in\Fca(t_m) \ \text{and}\ \tau_C > t_m \}
\end{align}
These \yo\ and \yoc\ are equivalent to \xo\ and \xoc\ because they see the same events, same filtrations, but have different measures, and agree on sets of measure zero \cite{shreve2004stochastic} Definition 1.6.3.
For example $ \Qbb_{\omega,C}$ are found by calibrating to the future $\Pbb_{\omega,C}$ measure observables at $t_m$ for each $\omega\in\Fca(t_m) \ \text{and}\ \tau_C > t_m$.

\subsubsection{Pricing at $t_0$}

Here we give the normal pricing, i.e. without \mb or \res.  This covers pricing with \res as this contract is priced in its entirety at $t_0$.

Derivative providers price \xva as the risk-neutral expected loss of a derivative, or portfolio, from counterparty default and the funding cost whilst the trade is alive.   We assume independence of exposure and default for simplicity.  Following \cite{burgard2014funding}, the \xva at inception is:
\begin{align}
\CVA(t_0;t_0,T) =& \LGD \int_{u=t_0}^{u=T}\lambda(u) e^{\int_{s=t_0}^{s=u}  -\lambda(s) ds}   \E^\QbbTz \left[ D_{r_F}(u)    \Pi^+(u)      \right]   du  \label{e:cva}\\
\FVA(t_0;t_0,T) =&  \int_{u=t_0}^{u=T}s_F(t) e^{\int_{s=t_0}^{s=u}  -\lambda(u) ds}   \E^\QbbTz \left[ D_{r_F}(u)    \Pi(u)      \right]   du  \label{e:fva}
\end{align}
$\CVA(t_0;t_0,T)$ means that the CVA is calculated at $t_0$ for exposure from $t_0$ to $T$ and similarly for FVA.  We make the definition
\begin{equation}
\xVA^\Qbb(t_0;t_0,T) := \CVA(t_0;t_0,T) + \FVA(t_0;t_0,T) \label{e:xva}
\end{equation}
where we include the measure that the \xva used for clarity.  Also:

$\lambda(t)$ = counterparty hazard rate.

$\Pi^+$ = positive exposure of position w.r.t. counterparty.

$r_F(t) := s_F(t) + r(t)$ = Bank funding cost, and separation into funding spread and riskless rate.

Since trades end on their \mb dates, \xva is calculated up to the \mb  date of the derivatives with \mbs, and to the full term with \ress, where $T$ is the date of the last payment.  To continue the trade after a \mb the client must enter a new trade and pay \xva on this continuation trade.

\subsubsection{Pricing with \mbs at $t > t_0$}

To price \mbs from the client point of view we need to price the trade and \xva after the \mb/\resg as well as the trade and \xva before the \mb/\resg.  
 
 Clients do not hedge their own default probability nor the funding cost of the provider so they value \xva in the real world, i.e. the \Pbb-measure.  Clients will only enter into a trade after a \mb if they survive so we need to consider this.

A key factor in \mb valuation is the setup of the continuation trade after the \mb.  Typically this will be at the money (ATM), not at the previous level.  The settlement at the \mb date provides the hedge against changes in riskless value from changes in market level.  This is the functional hedge aspect of the trade in action.

Assuming a single trade, without a \mb the \xva, here CVA and FVA, cost to a client is just  Equation \ref{e:xva}:
\begin{equation}
\xCCno(t_0; t_0, T)  = \xVA^\Qbb(t_0; t_0, T)  \label{e:xva2}
\end{equation}
The \res case is covered by the above when the exposures within Equations \ref{e:cva} and \ref{e:fva} are from the resetting trade.

With a \mb at $t_m$ the client cost is the sum of the \xva on the trade with the mandatory break, and the later continuation trade to original trade maturity
\begin{align}
\xCCwith(t_0, \omega; t_0, T)  =  \xVA^\Qbb(t_0; t_0, t_m)   +   \xVA^{\Qbb_{\omega,C}}(t_m; t_m,T)
\end{align}
$\xCCwith(t_0, t_m; t_0, T)$ is a random variable because it depends on the future state of the world via the events up to $t_m$, i.e. $\omega$, and the client survival  up to $t_m$ within $\Qbb_{\omega,C}$.  As we saw above, $\Qbb_{\omega,C}$ is a future risk-neutral measure dependent on earlier $\Pbb$ measures.  

The client cannot hedge $ \xVA^{\Qbb_{\omega,C}}(t_m; t_m,T)$ at $t_0$ with the street at a price the client will accept because the client considers that the observed CDS curve does not reflect the client's recovery post-crisis.  Also, counterparties may be reluctant to trade CDS referring to the client with the client.    In short, the client's view is that supply and demand for their CDS does not reflect future credit risk levels, but includes additional premia.   Another way of saying this is that the client does not calibrate the drift of their \Pbb\ measures to the current observed CDS curve.  

Below we look at examples of  how the \mb changes the total XVA cost to the client, $\MB(t_m, \omega)$,  as a function of the \mb date $t_m$ and the assumptions on recovery, i.e. $\Pbb_{\omega,C}$
\begin{align}
\MB(t_m, \omega) &:= \xCCno(t_0; t_0, T) - \xCCwith(t_0, \omega; t_0, T)   \\
	     &=   \xVA^\Qbb(t_0; t_0, T) - \left(      \xVA^\Qbb(t_0; t_0, t_m)   +   \xVA^{\Qbb_{\omega,C}}(t_m; t_m,T)      \right)
\end{align}
We characterize classes of $\omega$ by the change in credit spread of the client at $t_m$ relative to $t_0$.  

We now look at historical CDS shocks and recovery to inform the numerical examples.  

\section{Crises and recovery}

Here we analyze CDS shocks and their recovery.  The CDS universe used is pre-selected for a minimal level of liquidity, starts in May 2002 and ends May 2020.  The main indicator we use is the maximum of the 1Y and 5Y CDS spreads to allow for CDS curve and liquidity changes under stress.  

We want to detect shocks that are significant to firms and recoveries that are usable for hedging purposes, so data is prepared as follows to reduce effects of noise, insufficient data, and missing data.
\begin{itemize}
\item Only consider names from three regions, Asia, Europe, and North America, because these have the largest number of active names ($>$500 each).
\item Remove any name that has less than 2.1 years' data.  2.1 as a cutoff is derived from the window of 1 year for detecting shocks and the 1 year no-detect period after a shock detection.  Gaps are permitted and linearly interpolated.  We use a window size of one year so if there is less than two year's data the name will not provide a useful contribution.
\item Apply a 21-point median filter.  This takes the median across a month so that the results are not affected by daily noise.
\end{itemize}
Data preparation reduces the initial dataset from 10.1m observations to 6.6m and the total number of names from 5.4k to 3.4k.  Very roughly half of the names are active on any given date.    Obviously the results may be biased towards liquid names so this caveat should be included in making any use of the results in this paper.

We define a shock in historical CDS as:
\begin{itemize}
\item A shock for an individual company is an increase of CDS spread over a past window of at least a given size, where this shock occurs at least one window period after any previous shock.  
\begin{itemize}
\item The window period chosen is one year.  
\item We look at shocks sized 250bps, 500bps, and 1000bps.  Shock size is measured as
\begin{equation}
\text{shock\ size} := \text{CDS(t)} - \text{quantile}(10\%, \{\text{CDS}(u):\ t-1 \le u < t\})
\end{equation}
\end{itemize}
\item A crisis for the market is when the percentage of CDS names undergoing shocks is at least a given percentage of active CDS names.
\item Recovery is the change in CDS spread at fixed horizons after a shock for an individual company.  
\begin{itemize}
\item Change in CDS spread at 6m, 1y, 2y, 3y, 4y, and 5y horizons after each shock.  
\item CDS spread change at each horizon is defined as the change to the median CDS level at $\pm$5\%\ of the horizon.  This is to model clients having some flexibility on exactly when to transact any re-hedge, i.e. considering horizon $h$ with a shock date of $t$:
\end{itemize}
\end{itemize}
\begin{align}
\text{CDS\ spread\ change} :=& \text{quantile}(50\%, \{\text{CDS}(u):\ t + 0.95\times h \le u < t + 1.05\times h\}) \nonumber \\
  & {} -  \text{CDS(t)} \label{e:rec}
\end{align}

\begin{figure}
\begin{center}
\includegraphics[width=0.85\textwidth]{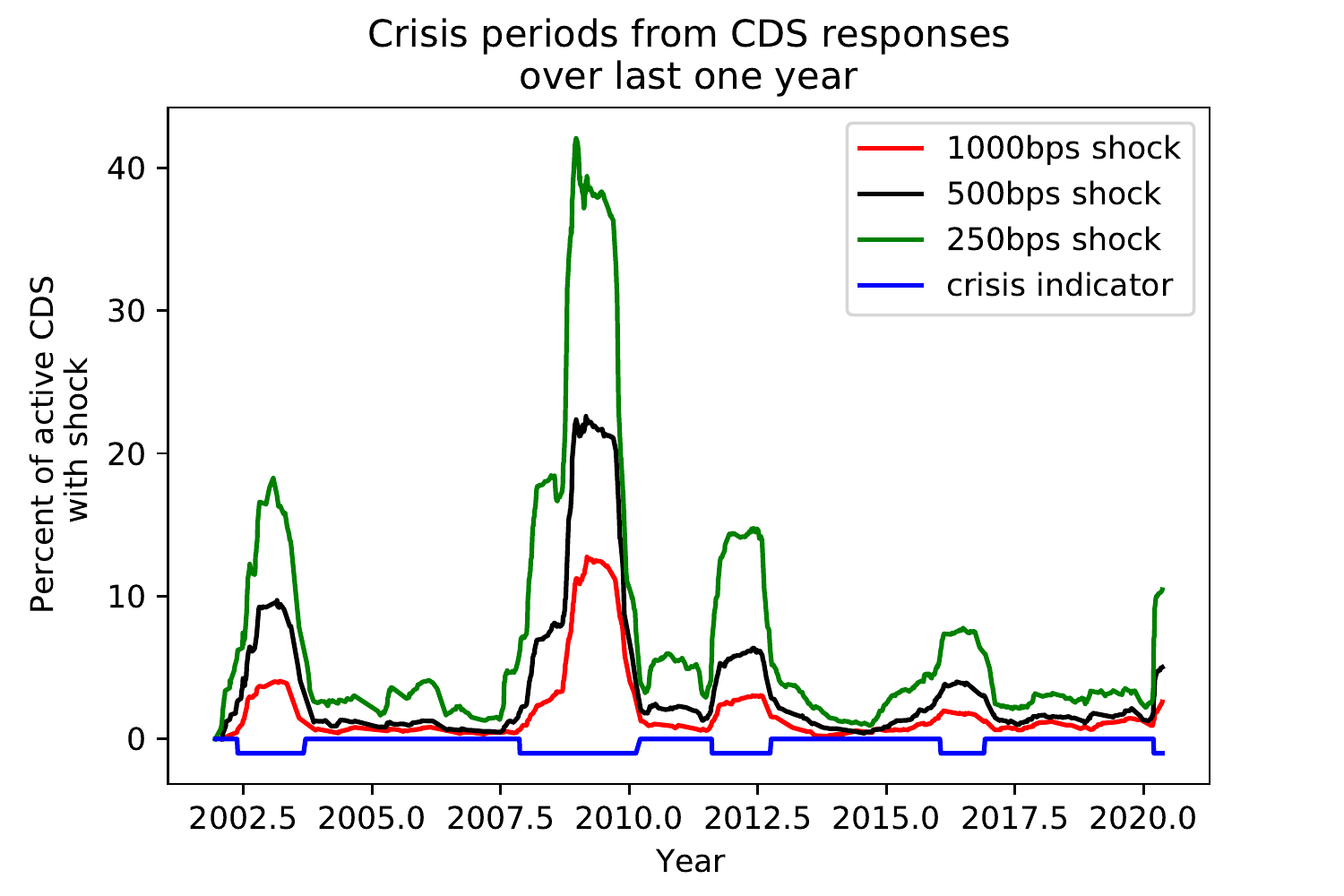}
\end{center}
\caption{Percent of active names with shocks over the last year.  Different colors correspond to different shock sizes: 250bps = green, 500bps = black, and 1000bps = red. Crisis periods are shown by lowered levels of the blue curve.  Small variations of this definition have no effect on results.}
\label{f:crisis}
\end{figure}

The blue line in Figure \ref{f:crisis} shows the definition of market crises used: 6\%\ of active names with at least a 250bps shock in the last one year.  This definition was chosen to highlight the periods with elevated percentages of CDS with shocks.  Small variations of this definition have little effect on results.

\section{Numerical results}

We first describe historical recovery from shocks and then give effects of alternative \xva management strategies.

\subsection{Recovery from shocks}

\begin{table}
\begin{center}
\begin{tabular}{lrrrrrrr}
\toprule
shock &  0.5y &  1.0y &  2.0y &  3.0y & 4.0y &  5.0y &    n \\
250.0 &       \multicolumn{6}{c}{horizon (years)}       &    at 2y   \\
\midrule
0.05  &  -920 & -1230 & -1490 & -1345 & -1500 & -1617 &  1686 \\
0.25  &  -171 &  -251 &  -316 &  -331 &  -337 &  -347 &  1686 \\
0.50  &   -28 &  -160 &  -205 &  -176 &  -212 &  -233 &  1686 \\
0.75  &   171 &    62 &   -91 &   -81 &   -93 &  -152 &  1686 \\
0.95  &  1331 &  1728 &   417 &   380 &   437 &   175 &  1686 \\
\bottomrule
\end{tabular}

\begin{tabular}{lrrrrrrr}
\toprule
shock &  0.5y &  1.0y &  2.0y &  3.0y &  4.0y & 5.0y &   n \\
500.0 &        \multicolumn{6}{c}{horizon (years)}       &    at 2y   \\
\midrule
0.05  & -1370 & -2293 & -2433 & -2277 & -2462 & -2769 &   898 \\
0.25  &  -342 &  -515 &  -613 &  -637 &  -691 &  -710 &   898 \\
0.50  &   -64 &  -350 &  -453 &  -438 &  -493 &  -534 &   898 \\
0.75  &   352 &    -9 &  -262 &  -266 &  -307 &  -416 &   898 \\
0.95  &  3140 &  2951 &   741 &   475 &   487 &   132 &   898 \\
\bottomrule
\end{tabular}

\begin{tabular}{lrrrrrrr}
\toprule
shock &  0.5y &  1.0y &  2.0y &  3.0y &  4.0y &  5.0y &    n \\
1000.0 &       \multicolumn{6}{c}{horizon (years)}       &    at 2y   \\
\midrule
0.05   & -1546 & -2750 & -3361 & -2943 & -3317 & -3026 &   469 \\
0.25   &  -715 & -1042 & -1241 & -1199 & -1280 & -1378 &   469 \\
0.50   &  -237 &  -812 &  -953 &  -915 &  -999 & -1096 &   469 \\
0.75   &   773 &   -89 &  -714 &  -658 &  -797 &  -936 &   469 \\
0.95   &  6557 &  5194 &   822 &   334 &   792 &  -221 &   469 \\
\bottomrule
\end{tabular}
\end{center}
\caption{Quantiles of distribution of changes in CDS spreads from  shocks for horizons of \{0.5y, 1y, 2y, 3y, 4y, 5y\} in crisis periods.  All shocks and changes are in bps.  The number of shocks in the last column (n) for the 2y horizon.  The first column gives the quantile of the distribution of the change in CDS spread.  We display \{5\%, 25\%, 50\%, 75\%, 95\%\} quantiles.}
\label{t:dist}
\end{table}

Table \ref{t:dist} gives the quantiles of distribution of changes in CDS spreads as defined in Equation \ref{e:rec} for horizons of \{0.5y, 1y, 2y, 3y, 4y, 5y\} in crisis periods.  We can observe that
\begin{itemize}
\item Looking at the median rows (0.50s) by two years most of the initial shock is recovered.  For the largest shock, 1000bps, 80\% of the recovery is after one year.
\item At least 5\%\ of the time there is no recovery.  Things get worse.
\item 25\%\ of the time there is mild recovery until five years when most of the shock is recovered.  For the largest shock, even in the 25th percentile 70\%\ of the recovery is present by two years.
\end{itemize}

There appears to be survivor bias in this analysis since we only observe CDSs that do not default.  However, from a \mb point of view this is correct because in the case of default the client is not concerned about trade renewal.  That is, we only want to consider cases where the client survives.  There is no bias from the \mb use and design perspective.

We now have a quantification of both recovery and risk or degree of recovery from historical CDS shocks.  Now we need to add the CVA quantification w.r.t. \mb and to bring the two parts together.

\subsection{Effects of \xva management strategies}

We now look at \xva management strategies informing the range of our analysis by the timescale of shock recovery, i.e. 1-5 years, in the previous section and the sizes of the observed shocks and recoveries, i.e. 250 to 1000bps.

We consider an example 10 year EUR IRS as of 2020-05-29, where the client receives the floating rate.  This is typical in that it provides the client with protection from increases in interest rates, and EUR is currently at historically low levels, although rates can go down as well as up beyond previous levels.

When pricing forward \xva we assume that the current interest curve and volatility is the same at the \mb point.  This assumption is often called same-as-now as opposed to risk-neutral where, for example, we would move up the yield curve.  We also consider changes in volatility at the \mb point  below.   We compare with using a \res which is priced at $t_0$ so cannot benefit from later changes of client credit risk but as mentioned above has the advantage of using conditional survival probability for the part of the trade after the reset (and all times in fact).

\subsubsection{\res}

Table \ref {t:xvaReset} shows the \xva reduction as a percentage of \xva charge without a \res for \res points at 1y to 5y and CDS shocks of 500bps and 1000bps.  Note that the CDS level is locked in for the whole life of the trade.  In this example the change in exposure from the different reset dates roughly balances the different default probabilities.  There is a 20\%\ to 25\%\ reduction in \xva for \res points at 1 to 5 years.  This reduction has little dependence on the CDS level.

Since the trade has a \res there is no dependence on the \Pbb\ measure, or later realized CDS levels or realized interest rate volatility levels.

\begin{table}
\begin{adjustwidth}{-0cm}{-0cm}
\begin{tabular}{llrrrrrrr}
\toprule
   &     &        & CDS level &     1 &     2 &     3 &     4 &     5 \\
IRS maturity & dVol & shock & reached &       \multicolumn{5}{c}{reset point (years)}        \\
\midrule
10 & 0.0 & 500.0  & 600.0  &  19.9 &  24.9 &  24.2 &  20.7 &  16.0 \\
   &     & 1000.0 & 1100.0 &  21.8 &  24.8 &  22.1 &  17.5 &  12.5 \\
\bottomrule
\end{tabular}
\end{adjustwidth}
\caption{\xva reduction as a percentage of \xva charge without a \res for \res points at 1y to 5y and CDS shocks of 500bps and 1000bps.  Note that the CDS level is locked in for the whole life of the trade.  dVol of zero means that there is no change to the interest rate volatility.}
\label{t:xvaReset}
\end{table}

\subsubsection{\mb and Restructuring}

We assume that the continuation trade is ATM.  Table \ref{t:xvaMB} shows the reduction in \xva compared to a trade without a \mb, or post-trade restructuring.  We assume that the restructuring rebate pays 100\%\ of the \xva and is available.  The continuation trade is at the future CDS level of the client, so we include a range of possibilities, including improvement and worsening.  Even with significantly worse CDS levels there is little increase in total \xva, less than 5\%.  For as-is CDS levels the \mb is roughly half as effective as a \res.  This is because the surviving client at the \mb date pays \xva without the benefit of the conditional survival probability: defaulting clients simply have no need of the continuation trade.

When the CDS level improves after the initial shock the reduction in \xva can be two to three times the reduction from a \res.  For a 500bps shock, starting from 100bps, the break-even w.r.t. a \res is roughly an improvement of 1/4 of the shock.  For a 1000bps shock the break-even is roughly 1/3 of the shock.  The \xva reduction pattern is almost always better with a shorter \mb date, provided the CDS level has improved.

\begin{table}
\begin{adjustwidth}{-0.5cm}{-0.5cm}
\begin{tabular}{lllrrrrrrr}
\toprule
   &     &        &      CDS level  &  &     1 &     2 &     3 &     4 &     5 \\
maturity & dVol & shock & reached & CDS change &       \multicolumn{5}{c}{mandatory break point (years)}       \\
\midrule
10 & 0.0 & 500.0  & 600.0  & -250.0  &  -4.4 &   2.0 &   4.9 &   5.9 &   5.7 \\
   &     &        &        &  0.0    &  12.6 &  15.9 &  15.7 &  13.8 &  11.1 \\
   &     &        &        &  125.0  &  23.5 &  24.5 &  22.2 &  18.5 &  14.2 \\
   &     &        &        &  250.0  &  36.4 &  34.5 &  29.7 &  23.8 &  17.6 \\
   &     &        &        &  500.0  &  69.4 &  59.4 &  47.7 &  36.0 &  25.4 \\
   &     & 1000.0 & 1100.0 & -500.0  &  -2.9 &  -3.1 &  -4.0 &  -4.2 &  -3.7 \\
   &     &        &        &  0.0    &   7.6 &   6.5 &   4.2 &   2.5 &   1.4 \\
   &     &        &        &  250.0  &  16.7 &  14.2 &  10.4 &   7.2 &   4.8 \\
   &     &        &        &  500.0  &  29.4 &  24.5 &  18.5 &  13.2 &   8.8 \\
   &     &        &        &  1000.0 &  71.8 &  57.0 &  42.3 &  29.7 &  19.5 \\
\bottomrule
\end{tabular}
\end{adjustwidth}
\caption{\xva reduction as a percentage of \xva charge without a \mb for \mb points at 1y to 5y and CDS shocks of 500bps and 1000bps. We also consider CDS change at the time of entering into the continuation trade.  Interest rate volatility and yield curve same-as-$t_0$ for continuation trade.  Negative reductions indicate increases.}
\label{t:xvaMB}
\end{table}

Table \ref{t:xvaMBvol} shows \xva reduction as a percentage of \xva charge without a \mb for \mb points at 2y for CDS shocks of 500bps and 1000bps. We also consider CDS change at the time of entering into the continuation trade.  Interest rate volatility differences covered are -10bps to +10bps, whilst the yield curve is same-as-$t_0$ for continuation trade.  We observe that there is significant interplay between the volatility effect and the CDS change effect as we would expect as both are important in \xva.  As the CDS recovery increases there is less relative effect of change in volatility.  

\begin{table}
\begin{adjustwidth}{-0.0cm}{-0.0cm}
\begin{tabular}{lllllrrr}
\toprule
   &        &        &   & dVol &  -10.0 &   0.0  &   10.0 \\
maturity & shock & reached & split & CDS change &       \multicolumn{3}{c}{volatility change (bps)}        \\
\midrule
10 & 500.0  & 600.0  & 2 & -250.0  &   18.8 &    2.0 &  -14.8 \\
   &        &        &   &  0.0    &   29.3 &   15.9 &    2.4 \\
   &        &        &   &  125.0  &   35.9 &   24.5 &   13.1 \\
   &        &        &   &  250.0  &   43.5 &   34.5 &   25.6 \\
   &        &        &   &  500.0  &   62.4 &   59.4 &   56.5 \\
   & 1000.0 & 1100.0 & 2 & -500.0  &   13.5 &   -3.1 &  -19.7 \\
   &        &        &   &  0.0    &   20.8 &    6.5 &   -7.8 \\
   &        &        &   &  250.0  &   26.7 &   14.2 &    1.6 \\
   &        &        &   &  500.0  &   34.6 &   24.5 &   14.5 \\
   &        &        &   &  1000.0 &   59.2 &   57.0 &   54.8 \\
\bottomrule
\end{tabular}
\end{adjustwidth}
\caption{\xva reduction as a percentage of \xva charge without a \mb for \mb points at 2y for CDS shocks of 500bps and 1000bps. We also consider CDS change at the time of entering into the continuation trade.  Interest rate volatility differences covered are -10bps to +10bps, whilst the yield curve is same-as-$t_0$ for continuation trade.  Negative reductions indicate increases.}
\label{t:xvaMBvol}
\end{table}

\FloatBarrier
\section{Discussion and Conclusions}

Here we have considered client \xva management using either \mbs / \resg or \ress as tools adapted for recovery from crises and normal times respectively, and the cross-over between them.  \resg are similar in \xva effects to \mbs but can be done on any date if the provider agrees and if an \xva rebate is given.   The issue when CDS levels are high is that a derivative locks in the client credit risk level and the provider's funding level on the trade date, for the life of the trade. 

Analysis of historical crises defined by CDS shocks 2002--2020 shows that recovery is largely complete two years after the initial shock considering the median CVA recovery.  For 500bps shocks the 75\%\ of the names recover by at least half by two years, with 5\%\ showing continuing deterioration.

We found that if the CDS level does not recover, or if there was no shock in the first place, then a \res for a 10y IRS is roughly twice as effective in reducing \xva as a \mb.  If the CDS level improves for the client by even 1/3 of the shock to the CDS level, then a \mb or restructuring is at least as good as a \res, and can be several times better.  Analysis of CDS shock recovery from historical crises indicates that this level of recovery occurs in at least 75\%\ of cases.

Pricing from the client point of view answers whether a \mb and then a continuation contract is a true break, i.e. two separate contracts, or just a single contract in practice.  For both parties the riskless price of the continuation trade after a \mb is different seen from the original start date compared to the continuation from a \res, because it is a \Qbb-in-\Pbb\ measure price not a \Qbb\ measure price.  We provided a precise definition of the relevant probability spaces and measures.   Also the client faces higher \xva with a \mb than with a single contract containing a \res. These differences are invisible when pricing from the usual bank point of view because then only the contract up to the \mb is priced.  However the client has to compare using a \res in a single trade, or a \mb/\resg with two sequential trades.

Hedge accounting is highly relevant and will be covered elsewhere in detail \cite{kenyon2020hedgeaccounting}.  A key aspect is that Accounting can follow the ``entity’s risk management objective and strategy for undertaking the hedge'' so is not limited to contracts that exist at some particular time, e.g. at original trade inception.  This objective and strategy requires ``formal documentation'' by the entity, see \cite{ifrs2018financial}, Section  6.4.1.b and must meet hedge effectiveness tests in Section B6.4.1 including effects of credit risk in Section B6.4.7.

This paper is almost unique in taking the client's perspective in \xva valuation using a real-world perspective, rather than considering valuation from the provider's side in the risk-neutral perspective.  However, consideration of \mb makes this a requirement as the provider is indifferent (all risk is hedged) whereas the client is exposed to changes in their own credit risk and the providers funding risk.  Since we are currently in the \cvs crisis as defined by CDS shocks we have considered \mb valuation from this point of view, i.e. within a crisis from the historical CDS analysis.  During normal times, or for clients unaffected by \xva, with no significant changes in CDS level a \res can be twice as effective as a \mb or restructuring.

\section{Acknowledgments}

The author would like to gratefully acknowledge discussions with Kohei Ueda, Hayato Iida, Robert Wendt, Susumu Higaki, Muneyoshi Horiguchi, Richard Kenyon.  and Dott. Donatella Barisani.

\bibliographystyle{chicago}
\bibliography{XVAbibliographyQ}

\end{document}